\documentclass{article}

\usepackage[utf8]{inputenc} 
\usepackage[T1]{fontenc}    
\usepackage{hyperref}       
\usepackage{url}            
\usepackage{booktabs}       
\usepackage{amsfonts,amsmath}       
\usepackage{nicefrac}       
\usepackage{microtype}      
\usepackage{lipsum}
\usepackage{graphics,graphicx,subfigure,url}
\usepackage{amsthm}

\usepackage{authblk}

\graphicspath{{figs/}}

\usepackage{fancyhdr}

\title{Liquid metal solves maze}

\author[$1$]{Andrew Adamatzky}
\author[$2$]{Alessandro Chiolerio}
\author[$3$]{Konrad Szaci{\l}owski}

\affil[$1$]{
  Unconventional Computing Laboratory, 
  Department of Computer Science and Creative Technologies, 
  University of the West of England,
  Bristol BS16 1QY UK}
\affil[$2$]{Center for Sustainable Future Technologies,
Istituto Italiano di Tecnologia, Via Livorno 60, 10144 Torino, Italy}
\affil[$3$]{Academic Centre for Materials and Nanotechnology, AGH University of Science and Technology, Mickiewicza 30, 30-059 Krak\'ow, Poland}

\begin{document}
\maketitle

\begin{abstract}
\noindent
A room temperature liquid metal features a melting point
around room temperature. We use liquid metal gallium due to its non-toxicity. A physical maze is a connected set of Euclidean domains separated by impassable walls. We demonstrate that a maze filled with sodium hydroxide solution is solved by a gallium droplet when direct current is applied between start and destination loci. During the maze solving the droplet stays compact due to its large surface tension, navigates along lines of the highest electrical current density due its high electrical conductivity, and goes around corners of the maze's corridors due to its high conformability. The  droplet maze solver has a long life-time due to the negligible vapour tension of liquid gallium and its corrosion resistance and its operation enables computational schemes based on liquid state devices. 

\vspace{3mm}
\end{abstract}


\section{Introduction}

Dynamic behaviour of liquid meta droplets has been known for a substantial amount of time. The first reports by Alessandro Volta and independently William Henry were published in 1800~\cite{Mollenkamp2019}. These initial reports concern phenomena related to the so-called "mercury beating heart", a fascinating experiment, in which a drop of metal pulsates changing its shape due to a sequence of oxidation and reduction processes taking part at its surface, which in turn result in changes of surface tension at water/metal interface. The full and comprehensive description of these processes was given recently by Najdolski et al~\cite{Najdolski}, however some mechanistic investigations have been reported much earlier \cite{Lin1974}.   Very recently electrochemical oscillators based on gallium droplets have been described in detail, including various oscillations modes \cite{Yu2018}. In 2019 Liu, Sheng and He proposed the concept of `liquid metal soft machines': implementation of actuation with liquid metal gallium and its alloys~\cite{liu2019}. Their extensive studies on shape changing~\cite{zhang2014synthetically}, electromagnetic propulsion~\cite{wang2015electromagnetic} and  self-propulsion~\cite{zhang2015self} demonstrated that gallium and similar alloys could be fruitful materials for the implementation of future liquid robots~\cite{chiolerio2017smart} and liquid computers~\cite{adamatzky2019brief}. Such cybernetic systems could be used to explore complex geometrically constrained spaces, where not other machines can enter. This new paradigm, of liquid state robots, could show particular advantages over specific application domains such as space, extreme environment exploration (where pressure, radiation, temperature are high enough to impair conventional solutions functionality), and as pointed out ultimately, also in high performance computation and data storage~\cite{tectomers}. Energetic systems able to store, transfer, harvest energy in the liquid state have also been conceived~\cite{CERES}. Mobility, as fundamental aspect of autonomy, after computation and homeostasis~\cite{homeostasis}, has been shown to provide an alternative way to transport information in liquid cellular automata~\cite{mobility}. To evaluate a potential of the `gallium robots' for programmable navigation in the constrained space we decided to test them on the task of a maze solving. 

To solve a maze means to find a route from the source site to the destination site in a geometrically constrained space. There are two scenarios of the maze problem: (1)~the solver does not know the whole structure of the maze 
and (2)~the solver knows the structure of the maze. Shannon's maze solving mechanical mouse Theseus~\cite{shannon1951presentation} and Wallace's maze solving computer~\cite{wallace1952maze} were used in the first scenario in the early 1950s. The second scenario maze solving employs the famous Lee algorithm\cite{lee1961algorithm, rubin1974lee} as follows. We start at the destination site. We label neighbours of the site with `1'. Then we label their neighbours with `2'.  Being at the site labelled $i$ we label its non-yet-labelled neighbours with $i+1$. Sites occupied by obstacles, or the maze walls, are not labelled. When all accessible sites are labelled the exploration task is completed. To extract the path from any given site of the maze till the destination site we start at the source site and then descend along the gradient of labels following the lowest numbers. In robotics the Lee algorithm was transformed into a potential method \index{potential method} pioneered in  \cite{pavlov1984method} and further developed in \cite{wyard1995potential, hwang1992potential}. The destination is assigned an infinite potential. Gradient is calculated locally. Streamlines from the source site to the destination site are calculated locally at each site by selecting locally the maximum gradient~\cite{connolly1990path}. 

The Lee maze solving algorithm is physically embedded in the mechanics of the experimental maze solvers proposed so far, see overview in ~\cite{adamatzky2017physical}. Examples of the experimental maze solvers are as follows. The shortest obstacle avoided path can be approximated by an electrical current and visualised with a glow-discharge~\cite{reyes2002glow, dubinov2014glow}, thermal camera~\cite{ayrinhac2014electric}, or as trace of 
self-assembled conductive particles~\cite{nair2015maze}. When a source of electricity is not available the shortest path can be found via Marangoni flow and then visualised with dyes~\cite{fuerstman2003solving} or droplets~\cite{lagzi2010maze,cejkova2014dynamics}. The path from the source to the destination can be explicitly presented by orientation of crystal needles during crystallisation in a supersaturated solution of sodium acetate trihydrate and water~\cite{adamatzky2009hot}. The path can be also approximated by gradients of a diffusion of chemo-attractants and then traced by living creatures interacting with such molecules, e.g. slime mould~\cite{adamatzky2012slime} or epithelial cells~\cite{scherber2012epithelial}. Finally the path can be calculated by analysing geometry of excitation waves in a thin layer of Belousov-Zhabotinsky medium~\cite{agladze1997finding,adamatzky2002collision}. 
Therefore is was so tempting to built an autonomous robotic entity capable of decision-making and information processing. The current study explores the possibility of creation of an autonomous self-propagating and environment-responsible liquid machine.

\section{Methods}
\label{methods}

\begin{figure}[!tbp] 
    \centering
    \includegraphics[width=\textwidth]{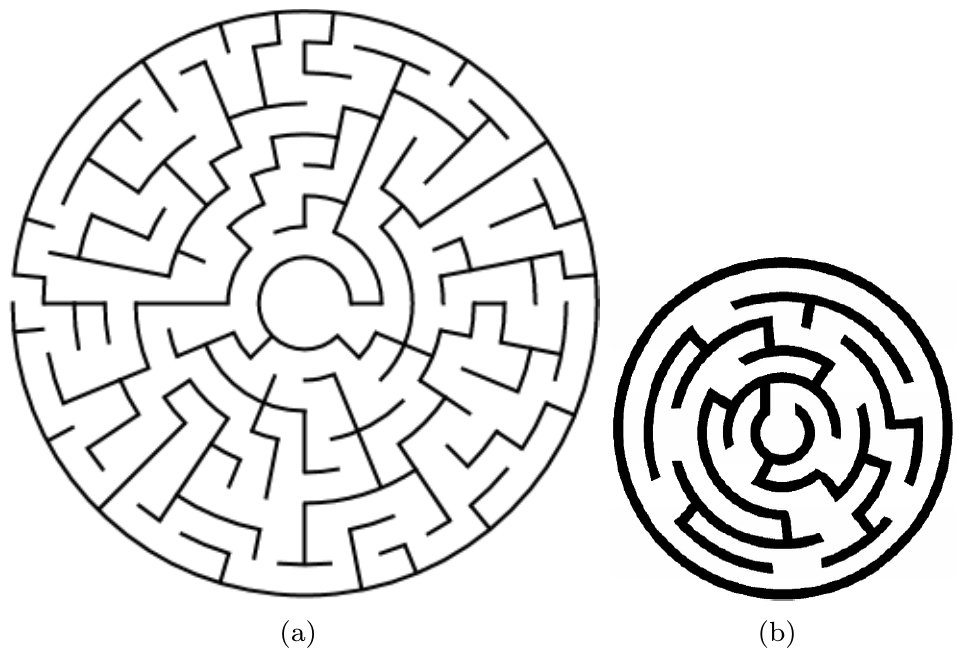}
      \caption{Topology of mazes used in experiments, scaling is 50\% of the original size.}
    \label{fig:mazes}
\end{figure}

Gallium (Sigma Aldrich, UK), its temperature is kept at 36\textsuperscript{o}C using a glass hot tray (RCHP-500G, Royal Catering, UK), two types of mazes were used (Fig.~\ref{fig:mazes}):
($M_1$)~PLA 3D printed maze 120~mm  diameter  with  6~mm  wide  and  10~mm  deep channels.
($M_2$)~nylon 70`mm  diameter  with  4~mm  wide  and 3~mm  deep  channels (Tesco’s  Toy  Mazes,  Tesco). The mazes were filled with 0.5~molar solution of NaOH (Sigma Aldrich, UK). 
Reasons for choosing these two mazes were (1)~to check whether the approach is scalable and (2)~to consider the case when Gallium droplet is fully immersed in the solution ($M_1$) and the case when it is half-immersed in the solution ($M_2$). The DC potential (30~V, 5~A) between starting point and destination sites was provided by  Tenma 72-10495  Bench Top Power Supply  via Platinum electrodes. The videos were recorded with Moto G6 and thermal images were done with FLIR ETS320. The Finite Element Method simulation was performed using Comsol COMSOL 5.3a, Licence No.13075366, d0946613b03f.

\section{Results}

\begin{figure}[!tbp] 
    \centering
 \includegraphics[width=\textwidth]{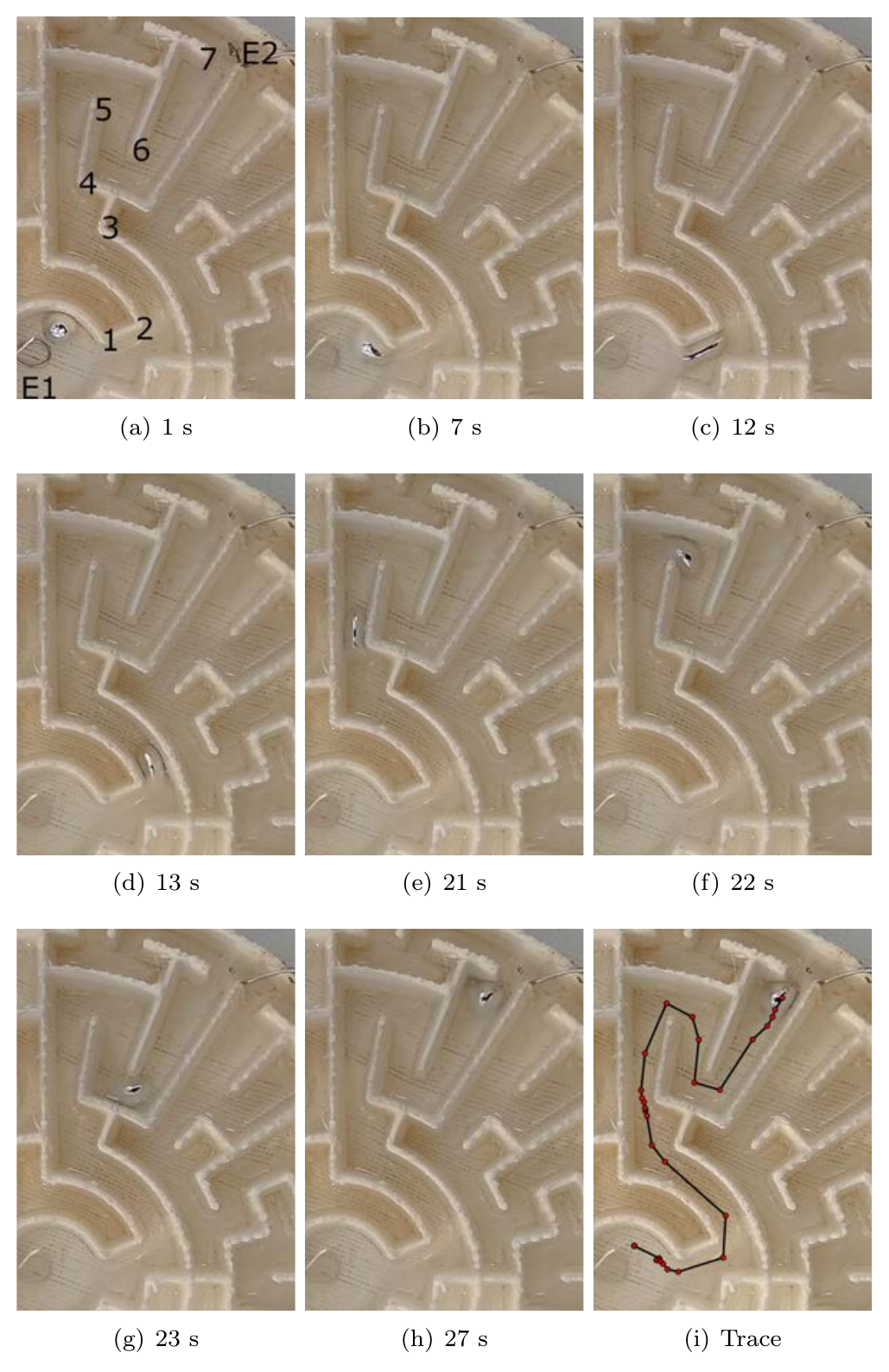}
    \caption{Snapshots of the gallium droplet travelling in the maze $M1$. Time from the start of recording is shown in captions. Trace of the liquid is shown in (h), where each dot corresponds to a position of the centre of the droplet during each frame of recording. Active $E1$ and passive $E2$ electrodes are marked, and waypoints are numbered in (a). See video in \cite{adamatzkyZenodoJuly2019}.}
    \label{fig:snapshotsLargeMaze}
\end{figure}

As soon as DC current is applied between positive electrode $E1$ (located at the start site in the central chamber of the maze) and negative electrode $E2$ (located at the destination site in the outside channel), Fig.~\ref{fig:snapshotsLargeMaze} the droplet jumps away from $E1$ by c. 3~mm but then remains motionless for few second (Fig.~\ref{fig:snapshotsLargeMaze}a). 

With continued application of DC some gas bubbles start forming at the site of the droplet close to $E1$, due to water electrolysis. In 7 s the droplet reaches turn 1  of the maze (Fig.~\ref{fig:snapshotsLargeMaze}b). While passing turn 1 the droplet got motionless for a few seconds at the wall connecting turns 1 and 2 (Fig.~\ref{fig:snapshotsLargeMaze}c) but then quickly dashes along corridor between 2 and 3  (Fig.~\ref{fig:snapshotsLargeMaze}d). Corners 3 and 4 almost do not affect the droplets motion (Fig.~\ref{fig:snapshotsLargeMaze}e) as it runs toward the turn 5. The droplet is delayed at the turn 5 by approximately 1~s (Fig.~\ref{fig:snapshotsLargeMaze}f) and at the turn 6 by approximately 4~s (Fig.~\ref{fig:snapshotsLargeMaze}g) till reaching the destination site in 27 s (Fig.~\ref{fig:snapshotsLargeMaze}h).

\begin{figure}[!tbp]
    \centering
 \includegraphics[width=\textwidth]{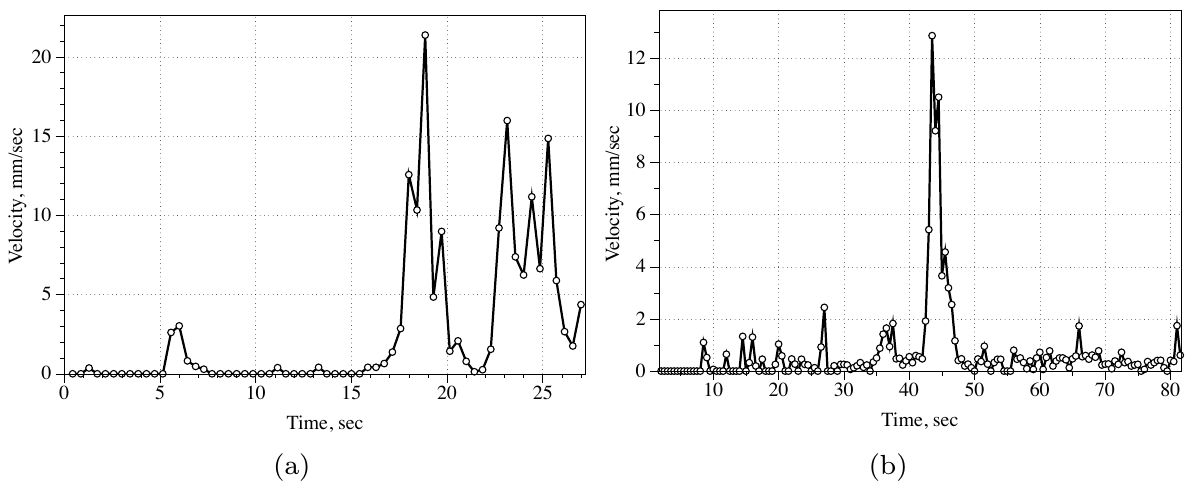}
    \caption{Dynamic of velocity of the gallium droplet travelling in (a)~maze $M_1$ (Fig.~\ref{fig:snapshotsLargeMaze}) and (b)~maze $M_2$ (Fig.~\ref{fig:snapshotsSmallMaze})}.
    \label{fig:speed}
\end{figure}

Tracing of the Ga droplet shown in Fig.~\ref{fig:snapshotsLargeMaze}i the positions of the droplet every 1~sec are marked. Their distribution, supported by the plot of the velocity (Fig.~\ref{fig:speed}a) covered by Ga each second of the experiment, indicate that the droplet is substantially delayed, or even often stays still while gathering a momentum,  when turning the corners. The delay length though does not seem to be affected by the angle of the turn.

\begin{figure}[!tbp]
    \centering
 \includegraphics[width=\textwidth]{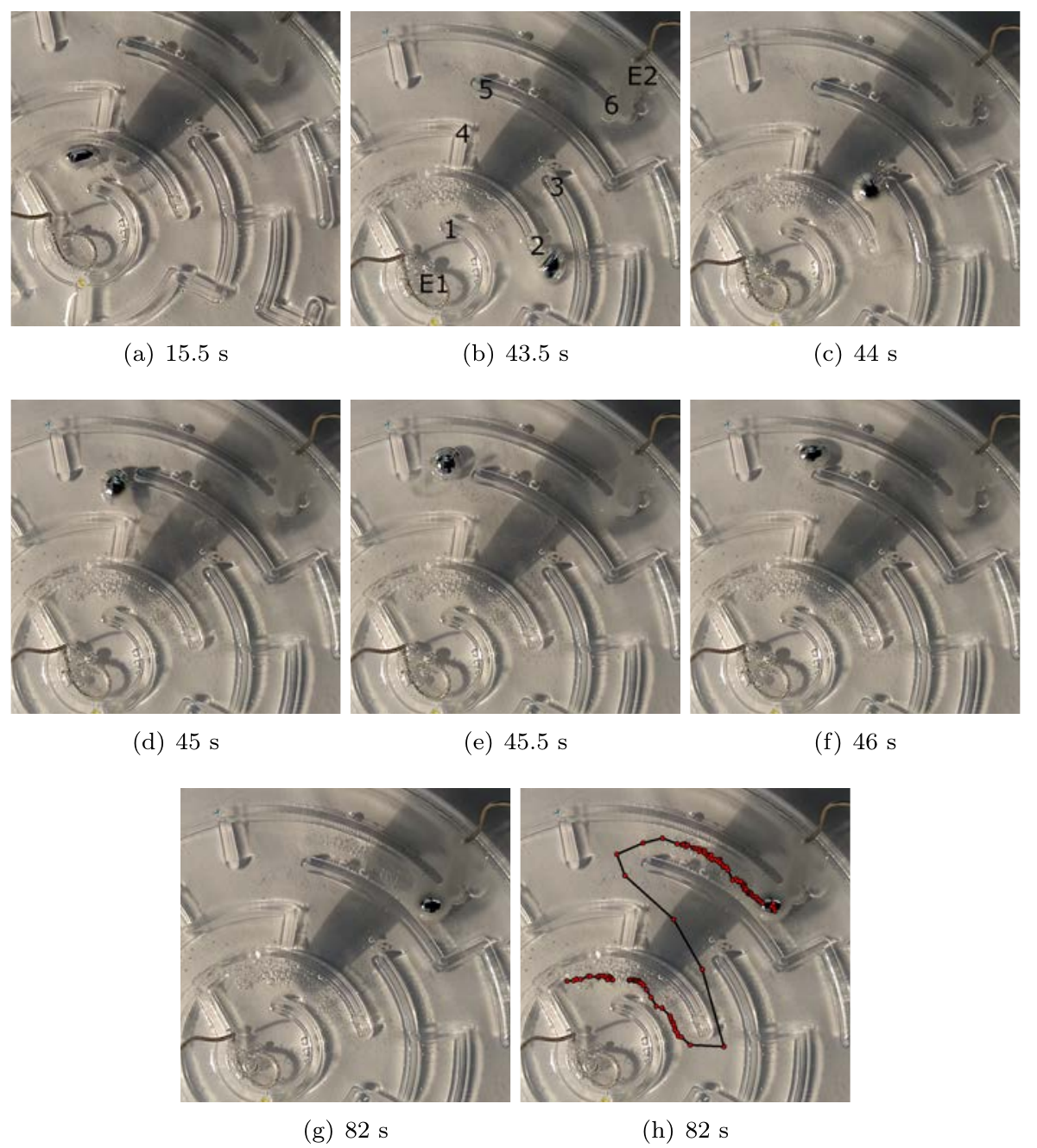}
    \caption{Snapshots of the gallium droplet travelling in the maze $M2$. Time from the start of recording is shown in captions. Trace of the liquid is shown in (h), where each dot corresponds to the position of the centre of the droplet during each frame of recording. Active $E1$ and passive $E2$ electrodes are marked, and waypoints are numbered in (b). See video in \cite{adamatzkyZenodoJuly2019}.}
    \label{fig:snapshotsSmallMaze}
\end{figure}

Navigation of a gallium droplet in maze $M_2$ is illustrated in Fig.~\ref{fig:snapshotsSmallMaze}. It takes approximately 82~s for the droplet to reach a vicinity of the electrode $E2$. Nearly half of this time is spent by the droplet near the starting electrode $E1$ (Fig.~\ref{fig:snapshotsSmallMaze}a) gradually accelerating. By approximately 43 s the droplet reaches the waypoint 2 (Fig.~\ref{fig:snapshotsSmallMaze}b). It then dashes by waypoints 3 (Fig.~\ref{fig:snapshotsSmallMaze}c), and 4 (Fig.~\ref{fig:snapshotsSmallMaze}d). The droplet reaches waypoint 5 by approximately 45 s (Fig.~\ref{fig:snapshotsSmallMaze}e). It takes the droplet approximately 1~sec to circumnavigate around the end of the wall, marked 5 in Fig.~\ref{fig:snapshotsSmallMaze}f. The passage from the waypoint 5 to the waypoint 6 poses a substantial difficulty for the droplet due to strong electrolyte convection flow resulting from vigorous evolution of gas bubbles at the electrode $E_2$. Yet by 82$^{nd}$ s since the beginning of the experiment the droplet approaches the electrode (Fig.~\ref{fig:snapshotsSmallMaze}g). Distribution of the droplet positions on the trace (Fig.~\ref{fig:snapshotsSmallMaze}h) and the velocity plot (Fig.~\ref{fig:speed}b) show that the slowest parts of trajectory are at the beginning of the journey, when the droplet gathers a momentum to overcome its static position and, and at the end of the journey, when the droplet struggles to overcome waves of bubbling from the target electrode. In all other parts the velocity is relatively stable, subject to minor perturbations. The surface of the maze $M_2$ is much smoother than that of the maze $M_1$ therefore the droplet spends less time (that the droplet in $M_2$) turning around corners.
The irregularities in the speed of motion of the gallium droplet may originate from irregularities of the oxide layer formed at the surface of gallium and consequently in small changes in the intensity of Marangoni flow which is directly responsible for droplet motion. The Marangoni flow results from local differences in surface tension of the droplet, as the case with some other types of travelling droplets~\cite{toyota2009self,baroud2010dynamics,nikolov2002superspreading}.  This also explains the induction period --- the droplet needs to accumulate sufficient amount of oxide (on the anodic side of the droplet) to start its motion.  

\begin{figure}[!tbp]
    \centering
 \includegraphics[width=\textwidth]{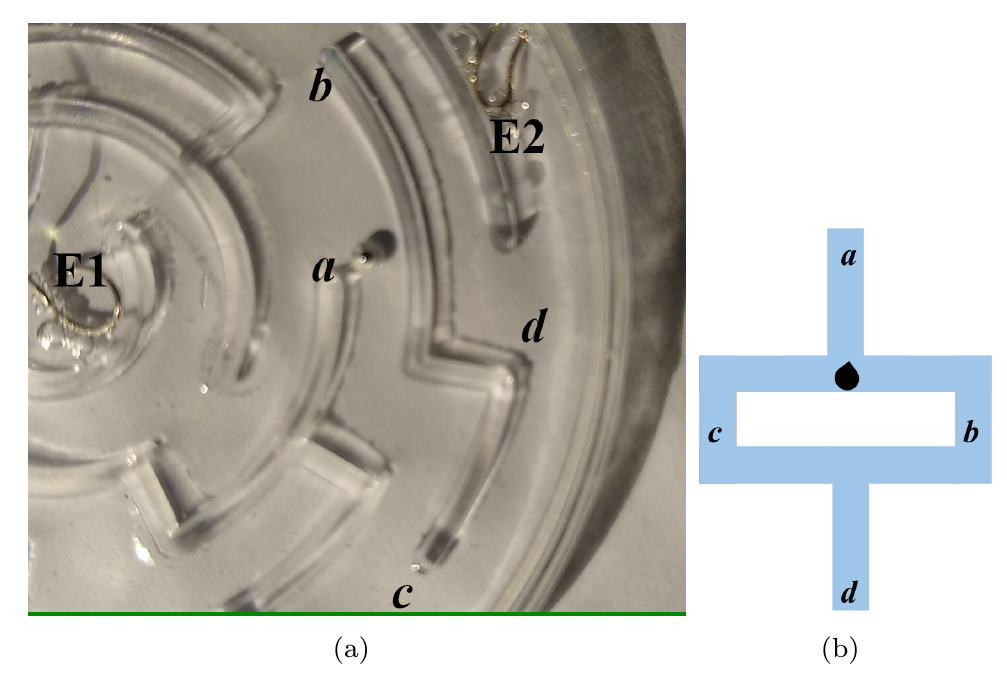}
    \caption{Illustration of the (a)~experiment and (b)~its scheme when the droplet was locked by two equal pathways of the current and thus unable to proceed further. As illustrated in the scheme (b): $|ab|+|bd|=|ac|+|cd|$. See also video in \cite{adamatzkyZenodoJuly2019}.}
    \label{fig:dropletlocked}
\end{figure}

In some experiments the droplet did not reach the target electrode $E_2$ due to its inability to choose between two branched pathways with nearly equal intensity of the current.  One of such experiments is illustrated in Fig.~\ref{fig:dropletlocked}. The droplet reaches site $a$. The length of pathways $p_{abd}$ and $p_{acd}$ is nearly equal: 38~mm and 42~mm in the maze $M_2$. Intensity of the current along a pathway is proportional to the length of the pathway. Therefore, on both sides of the droplet the intensity was similar. Thus, droplet was unable to choose which way to proceed (bifurcation point). 

\clearpage
\newpage 

\section{Discussion}

\begin{figure}[!tbp]
    \centering
 \includegraphics[width=\textwidth]{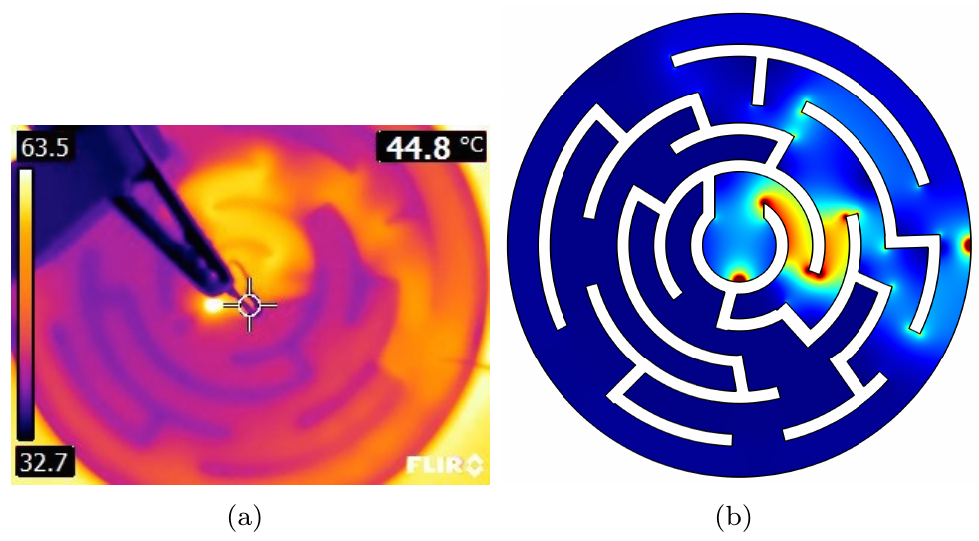}
    \caption{Visualisation of the highest current density in the maze $M_2$. Positive electrode is placed in the central chamber and negative electrode in the outside channel. (a)~Shortest path between central chamber and outside channel of the maze is seen as loci with highest temperature with thermal camera FLIR ETS320. (b)~Finite Element Method simulation showing in colour scale the current density distribution in the maze.}
    \label{fig:thermal}
\end{figure}

Why do gallium droplets move along the shortest path? This is because the intensity of electrical current is the strongest along the shortest path, as evidenced in Fig.~\ref{fig:thermal}, which is consistent with Kirchhoff's current law and Ohmic law, if we assume, for the sake of simplicity, that all the corridors in the maze have almost the same width. Then, the highest intensity current will be observed on a path with lowest resistivity (which corresponds to the shortest path if we assume identical specific resistivity of the liquid throughout the maze).  Approximation of the shortest collision-free path using electrical current was first demonstrated in a network of resistors in 1991~\cite{tarassenko1991analogue, tarassenko1991parallel}:  a space is represented as a resistor network, obstacles are insulators. Later visualisation of the path-solving with electrical current without physical discretisation of space was demonstrated in ~\cite{ayrinhac2014electric,ayrinhac2018electron}. The Finite Element Method (FEM) simulation performed in Comsol Multiphysics 5.3 environment over a 2D domain shows the distribution of current density in the maze $M_2$, where two electrodes have been set in a way similar to the experiments (outer circle, red spot at +180\textsuperscript{o}; inner circle, red spot at -90\textsuperscript{o}) and a voltage drop of 5V forced between them, while medium conductivity has been adjusted to NaOH solution as per experiments. The mesh has been built in order to avoid numerical deviations close to the edges, despite computational complexity. 

\begin{figure}[!tbp]
    \centering
 \includegraphics[width=\textwidth]{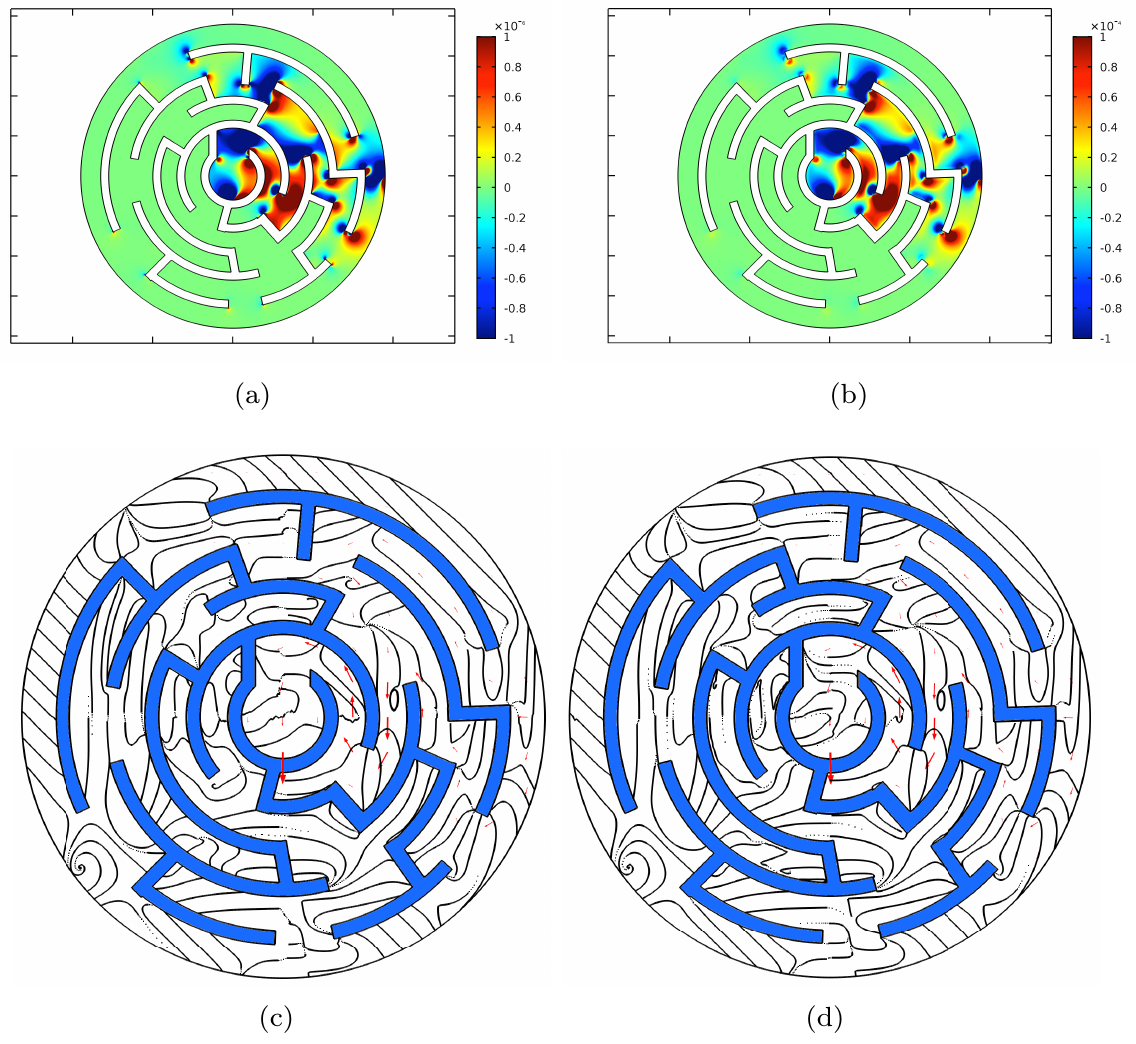}
    \caption{FEM results showing a comparison between \((dJ/dx, dJ/dy)\) quantity integrated over a rigid disk of size half the channel width (a), totally insulating walls and (b), insulating walls with Au conductive layer on each sharp edge; and \(J\) as a distribution of red arrows with the quantity \(grad J\) as black level curves (c), totally insulating walls and (d), insulating walls with Au conductive layer on each sharp edge.}
    \label{fig:electrical}
\end{figure}

This choice is fundamental in highlighting the role of wall conductivity, as shown in Fig.~\ref{fig:electrical}, a strategy that could be pursued to mitigate the high latency of Ga droplet during vertex bypasses. The top row (panels (a), totally insulating walls and (b), insulating walls with Au conductive layer on each sharp edge) shows in colour scale the \((dJ/dx, dJ/dy)\) quantity integrated over a rigid disk of size half the channel width, comparable to the size of Ga droplet. Therefore this quantity gives a measure of the complexity of the potential field that acts on the droplet as accelerating force. Integrating over smaller sizes does not change the geometry of this planar force field, but it affects its strenght in reason of a direct proportionality. One conclusion is that much smaller droplets will be subject to lower accelerating forces and a proper trade-off with viscous forces could result in faster motion. The bottom row (panels (c), totally insulating walls and (d), insulating walls with Au conductive layer on each sharp edge) shows \(J\) as a distribution of red arrows, clearly indicating the maze solving trajectory, and the quantity \(grad J\) as black level curves. Differences between the insulated maze case and the maze where sharp edges have been deposited with a gold electrically conductive shell suggest us that such a functionalisation, either of the edges or of the whole wall structure is effective in reducing the strength of potential localised at the edges, thus avoiding droplet latency/pinning.

 Further studies of the programmable navigation of liquid metal droplets will aim to answer the following questions. Given fixed electrical potential and current supplied what is a maximum distance between electrodes the droplet can propagate? Given a gallium droplet of a fixed size what is a narrowest gap the droplet can squeeze into? Will several gallium droplets co-interact, due to them affecting local conductivity of the medium, while navigating towards the exit of the maze?

\section{Acknowledgement}

AA acknowledges support of \mbox{EPSRC} with grant \mbox{EP/P016677/1}. 
Authors thank Dr. Tom Draper for kindly assisting with preparation of the NaOH solution. KS acknowledges support from National Science Centre (Poland) for financial support within the contract No. UMO-2015/18/A/ST4/00058.


\begin{thebibliography}{10}

\bibitem{adamatzky2009hot}
Andrew Adamatzky.
\newblock Hot ice computer.
\newblock {\em Physics Letters A}, 374(2):264--271, 2009.

\bibitem{adamatzky2012slime}
Andrew Adamatzky.
\newblock Slime mold solves maze in one pass, assisted by gradient of
  chemo-attractants.
\newblock {\em IEEE transactions on nanobioscience}, 11(2):131--134, 2012.

\bibitem{adamatzky2017physical}
Andrew Adamatzky.
\newblock Physical maze solvers. all twelve prototypes implement 1961 lee
  algorithm.
\newblock In {\em Emergent Computation}, pages 489--504. Springer, 2017.

\bibitem{adamatzky2019brief}
Andrew Adamatzky.
\newblock A brief history of liquid computers.
\newblock {\em Philosophical Transactions of the Royal Society B},
  374(1774):20180372, 2019.

\bibitem{adamatzkyZenodoJuly2019}
Andrew Adamatzky, Alessandro Chiolerio, and Konrad Szaci{\l}owski.
\newblock Supplementary materials. {L}iquid metal solves maze.
\newblock http://doi.org/10.5281/zenodo.3344993, 2019.

\bibitem{adamatzky2002collision}
Andrew Adamatzky and Benjamin de~Lacy~Costello.
\newblock Collision-free path planning in the belousov-zhabotinsky medium
  assisted by a cellular automaton.
\newblock {\em Naturwissenschaften}, 89(10):474--478, 2002.

\bibitem{agladze1997finding}
K~Agladze, N~Magome, R~Aliev, T~Yamaguchi, and K~Yoshikawa.
\newblock Finding the optimal path with the aid of chemical wave.
\newblock {\em Physica D: Nonlinear Phenomena}, 106(3-4):247--254, 1997.

\bibitem{ayrinhac2014electric}
Simon Ayrinhac.
\newblock Electric current solves mazes.
\newblock {\em Physics Education}, 49(4):443, 2014.

\bibitem{ayrinhac2018electron}
Simon Ayrinhac.
\newblock The electron in the maze.
\newblock In Andrew Adamatzky, editor, {\em Shortest Path Solvers. From
  Software to Wetware}, pages 409--420. Springer, 2018.

\bibitem{baroud2010dynamics}
Charles~N Baroud, Francois Gallaire, and R{\'e}mi Dangla.
\newblock Dynamics of microfluidic droplets.
\newblock {\em Lab on a Chip}, 10(16):2032--2045, 2010.

\bibitem{cejkova2014dynamics}
Jitka Cejkova, Matej Novak, Frantisek Stepanek, and Martin~M Hanczyc.
\newblock Dynamics of chemotactic droplets in salt concentration gradients.
\newblock {\em Langmuir}, 30(40):11937--11944, 2014.

\bibitem{tectomers}
Alessandro Chiolerio, Carsten Jost, Thomas~C. Draper, and Andrew Adamatzky.
\newblock The ph sensitivity of solvated tectomer electronics.
\newblock {\em arXiv preprint arXiv:1901.05519v1}, 2019.

\bibitem{chiolerio2017smart}
Alessandro Chiolerio and Marco~B Quadrelli.
\newblock Smart fluid systems: the advent of autonomous liquid robotics.
\newblock {\em Advanced Science}, 4(7):1700036, 2017.

\bibitem{CERES}
Alessandro Chiolerio and Marco~B. Quadrelli.
\newblock Colloidal energetic systems.
\newblock {\em Energy Technology}, 1800580:1--10, 2019.

\bibitem{connolly1990path}
Christopher~I Connolly, JB~Burns, and R~Weiss.
\newblock Path planning using {L}aplace's equation.
\newblock In {\em Robotics and Automation, 1990. Proceedings., 1990 IEEE
  International Conference on}, pages 2102--2106. IEEE, 1990.

\bibitem{dubinov2014glow}
Alexander~E Dubinov, Artem~N Maksimov, Maxim~S Mironenko, Nikolay~A Pylayev,
  and Victor~D Selemir.
\newblock Glow discharge based device for solving mazes.
\newblock {\em Physics of Plasmas}, 21(9):093503, 2014.

\bibitem{fuerstman2003solving}
Michael~J Fuerstman, Pascal Deschatelets, Ravi Kane, Alexander Schwartz,
  Paul~JA Kenis, John~M Deutch, and George~M Whitesides.
\newblock Solving mazes using microfluidic networks.
\newblock {\em Langmuir}, 19(11):4714--4722, 2003.

\bibitem{hwang1992potential}
Yong~K Hwang and Narendra Ahuja.
\newblock A potential field approach to path planning.
\newblock {\em Robotics and Automation, IEEE Transactions on}, 8(1):23--32,
  1992.

\bibitem{lagzi2010maze}
Istv{\'a}n Lagzi, Siowling Soh, Paul~J Wesson, Kevin~P Browne, and Bartosz~A
  Grzybowski.
\newblock Maze solving by chemotactic droplets.
\newblock {\em Journal of the American Chemical Society}, 132(4):1198--1199,
  2010.

\bibitem{lee1961algorithm}
Chin~Yang Lee.
\newblock An algorithm for path connections and its applications.
\newblock {\em IRE transactions on electronic computers}, 3:346--365, 1961.

\bibitem{Lin1974}
Shu-Wai Lin, Joel Keizer, Peter~A. Rock, and Herbert Stenschke.
\newblock On the mechanism of oscillations in the "mercury beating heart".
\newblock {\em Proc. Natl. Acad. Sci.}, 71(11):4477--4481, 1974.

\bibitem{liu2019}
Jing Liu, Lei Sheng, and Zhi-Zhu He, editors.
\newblock {\em Liquid metal soft machines}.
\newblock Springer, 2019.

\bibitem{Mollenkamp2019}
Hartwig Moellenkamp, Bolko Flintjer, and Jansen Walter.
\newblock 200 jahre "pulsierendes quecksilberherz".
\newblock {\em CHEMKON}, 1(2):117--125, 1995.

\bibitem{nair2015maze}
Aswathi Nair, Karthik Raghunandan, Vaddi Yaswant, Sreelal~S Pillai, and Sanjiv
  Sambandan.
\newblock Maze solving automatons for self-healing of open interconnects:
  Modular add-on for circuit boards.
\newblock {\em Applied Physics Letters}, 106(12):123103, 2015.

\bibitem{Najdolski}
Netodija Najdolski, Valentin Mirceski, Vladimir~M. Petrusevski, and Sani
  Demiri.
\newblock Mercury beating heart: Modifications to the classical demonstration.
\newblock {\em J. Chem. Educ.N}, 84(8):1292--1295, 2007.

\bibitem{nikolov2002superspreading}
Alex~D Nikolov, Darsh~T Wasan, Anoop Chengara, Kalman Koczo, George~A
  Policello, and Istvan Kolossvary.
\newblock Superspreading driven by marangoni flow.
\newblock {\em Advances in colloid and interface science}, 96(1-3):325--338,
  2002.

\bibitem{pavlov1984method}
VV~Pavlov and AN~Voronin.
\newblock The method of potential functions for coding constraints of the
  external space in an intelligent mobile robot.
\newblock {\em Soviet Automatic Control}, 17(6):45--51, 1984.

\bibitem{reyes2002glow}
Darwin~R Reyes, Moustafa~M Ghanem, George~M Whitesides, and Andreas Manz.
\newblock Glow discharge in microfluidic chips for visible analog computing.
\newblock {\em Lab on a Chip}, 2(2):113--116, 2002.

\bibitem{rubin1974lee}
Frank Rubin.
\newblock The lee path connection algorithm.
\newblock {\em IEEE Transactions on computers}, 100(9):907--914, 1974.

\bibitem{scherber2012epithelial}
Cally Scherber, Alexander~J Aranyosi, Birte Kulemann, Sarah~P Thayer, Mehmet
  Toner, Othon Iliopoulos, and Daniel Irimia.
\newblock Epithelial cell guidance by self-generated egf gradients.
\newblock {\em Integrative Biology}, 4(3):259--269, 2012.

\bibitem{shannon1951presentation}
Claude~E Shannon.
\newblock Presentation of a maze-solving machine.
\newblock In {\em 8th Conf. of the Josiah Macy Jr. Found.(Cybernetics)}, pages
  173--180, 1951.

\bibitem{tarassenko1991analogue}
L~Tarassenko and A~Blake.
\newblock Analogue computation of collision-free paths.
\newblock In {\em Robotics and Automation, 1991. Proceedings., 1991 IEEE
  International Conference on}, pages 540--545. IEEE, 1991.

\bibitem{tarassenko1991parallel}
Lionel Tarassenko, Gillian Marshall, Felipe Gomez-Castaneda, and Alan Murray.
\newblock Parallel analogue computation for real-time path planning.
\newblock In {\em VLSI for Artificial Intelligence and Neural Networks}, pages
  93--99. Springer, 1991.

\bibitem{toyota2009self}
Taro Toyota, Naoto Maru, Martin~M Hanczyc, Takashi Ikegami, and Tadashi
  Sugawara.
\newblock Self-propelled oil droplets consuming “fuel” surfactant.
\newblock {\em Journal of the American Chemical Society}, 131(14):5012--5013,
  2009.

\bibitem{homeostasis}
Scott~J. Turner.
\newblock Homeostasis as a fundamental principle for a coherent theory of
  brains.
\newblock {\em Philosophical Transactions R. Soc. B}, 374(20180373), 2019.

\bibitem{mobility}
William~F. Vining, Fernando Esponda, Melanie~E. Moses, and Stephanie Forrest.
\newblock How does mobility help distributed systems compute?
\newblock {\em Philosophical Transactions R. Soc. B}, 374(20180375), 2019.

\bibitem{wallace1952maze}
Richard~A Wallace.
\newblock The maze solving computer.
\newblock In {\em Proceedings of the 1952 ACM national meeting (Pittsburgh)},
  pages 119--125. ACM, 1952.

\bibitem{wang2015electromagnetic}
Lei Wang and Jing Liu.
\newblock Electromagnetic rotation of a liquid metal sphere or pool within a
  solution.
\newblock {\em Proceedings of the Royal Society A: Mathematical, Physical and
  Engineering Sciences}, 471(2178):20150177, 2015.

\bibitem{wyard1995potential}
L~Wyard-Scott and Q-HM Meng.
\newblock A potential maze solving algorithm for a micromouse robot.
\newblock In {\em Communications, Computers, and Signal Processing, 1995.
  Proceedings., IEEE Pacific Rim Conference on}, pages 614--618. IEEE, 1995.

\bibitem{Yu2018}
Zhenwei Yu, Frank~W. Yun, David Cortie, Lei Jiang, and Xiaolin Wang.
\newblock Discovery of a voltage-stimulated heartbeat effect in droplets of
  liquid gallium.
\newblock {\em Phys. rev. Lett.}, 121:024302, 2018.

\bibitem{zhang2014synthetically}
Jie Zhang, Lei Sheng, and Jing Liu.
\newblock Synthetically chemical-electrical mechanism for controlling large
  scale reversible deformation of liquid metal objects.
\newblock {\em Scientific reports}, 4:7116, 2014.

\bibitem{zhang2015self}
Jie Zhang, Youyou Yao, Lei Sheng, and Jing Liu.
\newblock Self-fueled biomimetic liquid metal mollusk.
\newblock {\em Advanced Materials}, 27(16):2648--2655, 2015.

\end{thebibliography}
\end{document}